# Multiple Frequency Selection in DVFS-Enabled Processors to Minimize Energy Consumption

*Nikzad Babaii Rizvandi[1,2], Albert Y. Zomaya[1], Young Choon Lee[1], Ali Javadzadeh Boloori[1,2], Javid Taheri[1]*

[1] Center for Distributed and High Performance Computing
School of Information Technologies, University of Sydney
NSW 2006, Australia

[2] National ICT Australia (NICTA), Australian Technology Park
Sydney, NSW 1430, Australia
{nikzad, yclee}@it.usyd.edu.au
{albert.zomaya, Javid.taheri}@sydney.edu.au
ajav4801@uni.sydney.edu.au

## 1. INTRODUCTION

Research on low power systems has received a great amount of attention in recent years since the sustainability of current technologies and practices has become a serious issue. A few example systems where lowering power usage is critical are:

- *Wireless sensors:* several sensors extract data from the environment concurrently, transmit these data to a processing unit and receive processed data accompanied by appropriate commands from the processing unit [1-4]. The sensors and their receiver/transmitter are generally powered by battery and/or solar cells.

- *Satellite circuits:* Satellites typically involve massive number of complex circuits that must work in low power. These circuits are supplied by solar cells, the only available power supply in satellites.
- *Robots and surveillance devices:* these devices are heavily used in army, mine extraction and in difficult or unsafe environments for humans.
- *Cell phones and laptops:* these devices are powered by batteries which are expected to work for a long time.

In the meantime, stiff increases in energy price and the environmental impact of carbon dioxide emissions associated with energy generation and transportation have forced the issue of reducing energy consumption to be extended to a broader range of system including High Performance Computing Systems (HPCS).

Various issues such as resource management in both software and hardware levels must be addressed to reduce energy consumption in HPCS. An important issue in hardware resource management is how to reduce power usage in processors. In the recent past, many hardware-based approaches have been made to efficiently reduce energy consumption, particularly for processors. Dynamic voltage-frequency scaling (DVFS) is perhaps the most appealing method incorporated into many recent processors. Energy savings with this method is based on the fact that the power consumption in CMOS circuits has direct relation with frequency and the square of voltage supply. In this case, the execution time and power consumption can be controlled by switching between processor's frequencies and voltages. Although this approach was initially designed for single processor task scheduling [5], it has recently received much attention in multiprocessor systems as well [6, 7].

DVFS technique and task scheduling can be combined in two ways: (1) schedule generation, and (2) slack reclamation. In the schedule generation, tasks graph are (re)scheduled on DVFS-enabled processors in a global cost function including both energy saving and makespan to meet both energy and time constraints at the same time [8, 9]. In slack reclamation, which works as post processing procedure on the output of scheduling algorithms, DVFS technique is used to minimize the energy consumption of tasks in a schedule generated by a separate scheduler. The existing methods based on DVFS technique, however, have two major shortcomings: (1) most of them focus on schedule generation and do not adequately take the slack reclamation approaches into account to save more energy, and (2) the existing slack reclamation methods use only one frequency for each task among all discrete set of processor's frequencies. Using one frequency usually results in uncovered slack time where processor and other devices only waste energy.

In this chapter we focus on slack reclamation and propose a new slack reclamation technique, Multiple Frequency Selection DVFS (MFS-DVFS). The key idea is to execute each task with a linear combination of more than one frequency such that this combination results in using the lowest energy by covering the whole slack time of the task. We have tested our algorithm with both random and real-world application task graphs and compared with the results in previous researches in [7] and [10]. The experimental results show that our approach can achieve energy almost identical to the optimum energy saving.

# 2. Energy Efficiency in HPCSs

Many of electronic systems in our life such as satellite systems, cell-phones, game instruments and so on are using rechargeable batteries as their power supplies. Although the battery capacity has been grown significantly in recent years (the battery capacity increases 5% per year), battery life is still the major drawback for most of electronic systems. In addition to power-aware battery-based systems, the issue of energy consumption has recently attracted a great amount of attention in high performance computing systems (HPCS). Energy consumption issue in such systems can be classified into three groups: (1) system-level resource allocation, (2) service-level energy-load distribution, and (3) task scheduling level (Figure 1).

In the system-level, the problem is how to distribute computational resources (e.g. CPU, network, memory and I/O) between large scale data storages and processing centers (such as supercomputers and data centers). Fairly distribute resources among applications (or services) not only requires to obtain individual adaptation among resources but also needs to understand the interaction between individual resources when they work as a system. Therefore, the big challenge here is to find both the relationship among system resources and their trade-off, which may cause an optimal balance between performance, QoS and energy consumption [11]. Among different technologies in system-level for managing resources between workloads, virtualization becomes a key technology in data centers. Virtualization allows the computational resources to be shared between different workloads. Many of incoming workloads to data centers are medium size workloads which often require a small fraction of the computational resources. The servers typically spend around 70% of their maximum power consumption even in low utilization. With virtualization, such workloads can be run within a virtual machine (VM) causing significant saving in overall energy usage. The associated VMs may require fewer amounts of resources and therefore they can be run on a single hardware unit. It is obvious that less hardware is used in overall, less energy is wasted for both working on and cooling of the servers.

In the service-level, energy reduction by load balancing, scheduling and mapping workloads is concerned. The main challenge is to utilize appropriate algorithms to both multiplex/demultiplex workloads in order to save energy and make a trade-off between performance and service cost reduction because of energy savings. Also, to avoid hotspot in data centers due to high-loaded nodes, services can be moved from nodes with high-load and high temperature to nodes with smaller load and lower temperature. Generally, this movement of services should happen when the destination nodes can operate the services in an energy efficient way [11].

In site-level data/task scheduling, the focus of this chapter, the operating system (OS) and hardware configuration such as dynamic power management, micro-architecture techniques and dynamic voltage scaling are used to decrease power. Here, the typical question could be:

*"What is the suitable OS/hardware configuration to process tasks in the shortest possible time and with minimum energy?"*

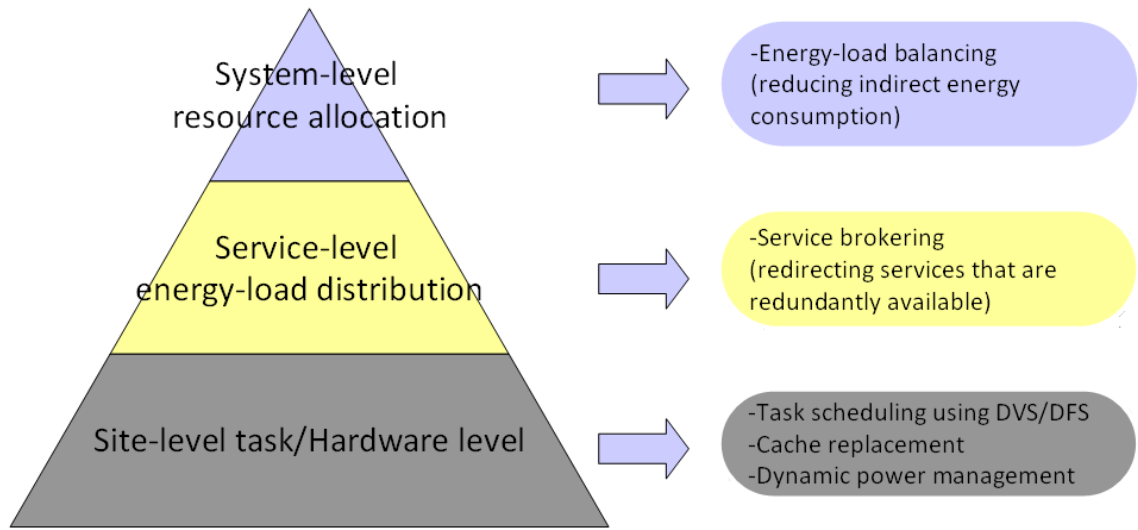

Figure 1. Energy consumption levels in HPCS

## 3. Exploitation of dynamic voltage-frequency scaling

Dynamic voltage-frequency scaling is a modern technique in computer architecture to reduce the energy consumption of microprocessors or control the amount of the generated heat by the circuit. This technique is commonly utilized in battery-based devices such as laptops and cell phones where decreasing the energy usage of battery is necessary. In addition, DVFS is used in high-computing nodes not only to decrease the power of the nodes but also to save more energy to cool down the nodes' places. An approximation model shows that the dynamic power in CMOS circuits is a linear function of both switching frequency and voltage square as: $C.V^2.f$, where C is the effective switching capacity per clock cycle. Therefore, a workload (or task) can save more energy when it is executed in lower voltage and frequency. In general, a computing node executes several tasks with inter-task relationships (e.g., precedence constraints) simultaneously. These inter-task relationships typically incur slack time (idle time) between tasks where can be used by DVFS to reduce energy usage. Specifically, the slack time associated with a task is utilized to execute the task in a lower voltage-frequency; this in turn results in energy reduction.

There are two ways to combine scheduling and Dynamic Voltage-Frequency Scaling: (1) independent slack reclamation, and (2) integrated scheduling generation. The existing methods in literature based on these combinations have two major limitations: (1) most of them focus on integrating DVFS and scheduling (integrated schedule generation) and do not sufficiently consider the slack reclamation approaches to save more energy, and (2) the existing slack reclamation methods use only one frequency for each task among all discrete set of processor's frequencies. Using one frequency usually results in uncovered slack time where processor and other devices only waste energy.

## *3.1. Independent slack reclamation*

Independent slack reclamation, works on the output of other scheduling algorithms as a post processing procedure by applying DVFS technique to minimize energy consumption of generated tasks by a scheduler. In [7] Kimura et al proposed an energy reduction algorithm for power-scalable clusters supporting DVFS. In a simplified version of this algorithm, the appropriate frequency is chosen among a set of processor's frequencies for each task regarding its slack time. Another algorithm was proposed in [10] to reclaim slack time for each task in a DAG by linear combination of the processor highest and lowest frequencies. To the best of our knowledge, among existing energy-aware algorithms in HPCS, these two methods are the most similar approaches to our MFS-DVFS algorithm presented in this chapter. We address the simplified version of these two algorithms as Reference DVFS (RDVFS) and Maximum-Minimum-Frequency DVFS

(MMF-DVF) in the rest of this chapter and will use them as benchmarks to evaluate the performance of our proposed algorithm.

## *3.2. Integrated scheduling generation*

In integrated schedule generation, tasks graph are (re)scheduled on DVFS-enabled processors using a global cost function including both energy saving and makespan to meet both energy and time constraints at the same time [8, 9]. Therefore, the final scheduling will be a trade off between makespan and energy. Kappiah et al in [12] presented Just-in-time DVFS technique to fill slack time in MPI programs. They utilized a system called Jitter to reduce the frequency on nodes with more slack times and fewer computations. Jitter aimed to make sure that the tasks came just in time without increasing overall execution time. DVS technique was applied in [8] on processors that did not work in peak performance during execution of a parallel application. The best processor frequency of each task was selected by analyzing computation and communication power profiles collected prior to the execution. A method to reduce power consumption was presented in [13] by adaptively activating and deactivating hardware resources and in particular, memory for intensive HPC applications. Cache missing in accessing the main memory also plays an important role in adjusting and triggering processors slack times. Lee and Zomaya in [9] presented a DVFS-based algorithm to minimize both completion time and energy consumption of precedence-constrained parallel jobs on HPC systems. This method tried to minimize a summation of two cost functions: completion time and energy. Consequently, the final result was a trade-off between the quality of scheduling and energy consumption. The concept of energy scalability in formal terms was introduced by Ding et al. in [14]. In addition to studying energy efficiency/iso-efficiency concept, they extended an analytical model to investigate the tradeoff between performance and energy saving in HPCS. Molnos et al in [15] classified the slack times in real-time applications into static, work and shared lack groups for multiple dependent tasks on multiple DVFS-enabled processors. They proposed a dynamic dependency-aware task scheduling to adjust voltage/frequency of each processor regarding tasks' real time deadlines. A profiled-based power-performance optimization method was presented in [16] to also utilize DVFS in HPCS. Here, the execution of a program was divided into several regions. In trial steps, profile information of each region, including power and execution profiles was extracted and then utilized to find its best combination of processors' voltages and frequencies. In [17], an upper limit for system energy usage was selected externally. Subsequently, a combination of performance modeling and performance prediction was applied to reduce execution times with respect to their predefined energy usage upper limit. After creating models for both execution time and energy consumption, key parameters of models were estimated by executing a program for a small number of times and then regressing the estimated parameters. Here, for better estimation of parameters, the following steps were iterated until a proper schedule is achieved: (1) using models to predict each possible scheduling of tasks, (2) executing the program a few times with the best predicted schedule and (3) updating estimated key parameters. Rountree et al in [18] proposed an energy-aware schedule generation algorithm for DVFS-enabled processors where a combination of all processor frequencies is involved into an overall linear programming optimization.

# 4. Preliminaries

In this section, the system, application and energy models used in our study have been described.

## *4.1. System and application models*

In this work, we assume an HPC system comprising of N homogeneous processors with individual memories. The switching time from one frequency to another is typically in microseconds (between $30\mu\sec$ and $150\mu\sec$ refer to [19]) while the execution time of tasks is in milliseconds. Therefore, compared with tasks' execution time, the switching time can be ignored. We consider a set of *M* dependent tasks denoted as $A^{(1)}, A^{(2)},..., A^{(M)}$ represented by task graph or directed acyclic graph (DAG). The k[th]-task ($A^{(k)}$) have the following parameters (Figure.2-a): (a) $t_{OS}^{(k)}$ is the task execution time in the original scheduling without slack reclamation, (b) $T^{(k)}$ is the whole time the processor assigns to this task. This time is a summation of the task's execution and slack times, (c) $t_i^{(k)}$ represents the task execution time when it is executed in frequency $f_i$, and (d) $K^{(k)}$ is the number of tick cycles required for executing this task. This parameter can be calculated as: $K^{(k)} = f_N t_{OS}^{(k)} = f_{RD}^{(k)} t_{RD}^{(k)}$, where $f_N$ is the highest processor frequency, and (e) $f_{RD}^{(k)}$ and $t_{RD}^{(k)}$ are frequencies calculated from RDVFS algorithm, explained in section 5.2 and its associated time, respectively.

## 4.2. Energy model

A typical DVFS-enabled processor can execute a task in a discrete set of frequencies $(f_1 < f_2 < ... < f_{N-1} < f_N)$. For example, AMD Turion MT-34 can operate at six frequencies ranging from 800MHz to 1800MHz [5]. The power consumption of a processor consists of two parts: (1) dynamic part that is mainly related to CMOS circuit switching energy, and (2) static part that addresses the CMOS circuit leakage power [20]. In CPUs, the power consumption is formulated as [21]:

$$\begin{cases} P_{dynamic} = C_{eff} f v^2 \\ P_{leakage} \propto v \end{cases} \quad (1)$$

Here, $C_{eff}$, $f$ and $v$ represent the effective capacitance, processor's frequency and voltage, respectively. Because the leakage power is always negligible compared with the dynamic power [20], the overall energy consumption of $k^{th}$-task $(A^{(k)})$ in DAG is calculated as:

$$E^{(k)} = P_{dynamic} t_i^{(k)} + P_{Idle}(T^{(k)} - t_i^{(k)}) \quad (2)$$

CPU power consumption can be modeled as a convex function of frequency as $P_{dynamic} = \alpha f^3 + \gamma$ [21]; Therefore, the energy of $k^{th}$-task $(A^{(k)})$ in Eqn.2 is changed to:

$$E^{(k)} \approx (\alpha f^3 + \gamma) t_i^{(k)} + P_{Idle}(T^{(k)} - t_i^{(k)}) \quad (3)$$

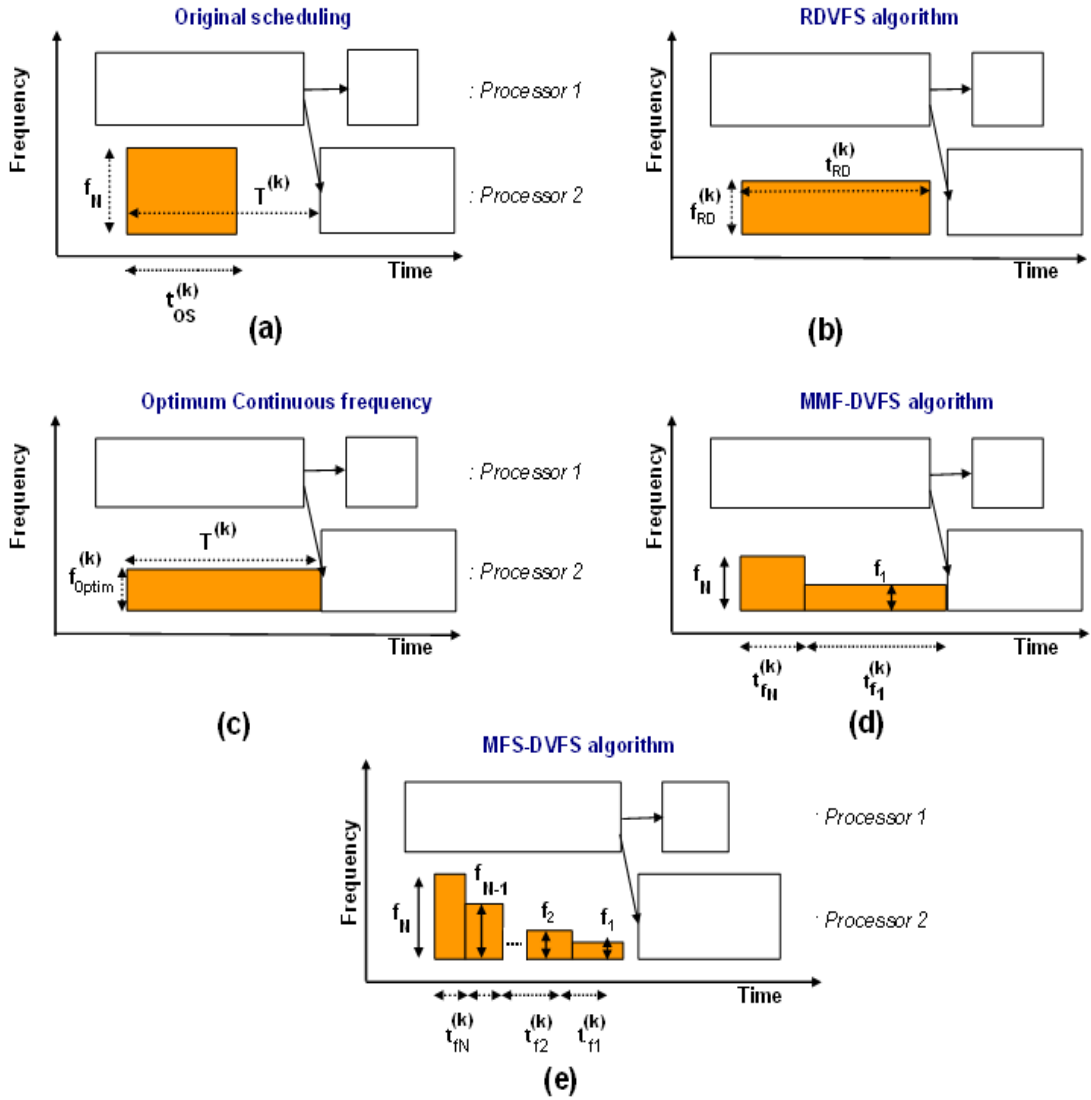

Figure 2. Time representation of MFS-DVFS and other algorithms: (a) The original scheduling, (b) The RDVFS algorithm, (c) the optimum continuous frequency, (d) the MMF-DVFS algorithm and (e) our proposed method in this chapter (MFS-DVFS).

# 5. Energy-aware scheduling using DVFS

In this section, we explain existing DVFS-based approaches to reduce energy consumption of processors by reclaiming the slack time for each task. In the end, we present our algorithm, MFS-DVFS, that uses a linear combination of frequencies to solve the stated problem.

## *5.1. Optimum Continuous Frequency*

The optimal approach to remove slack time and as a result, reduce energy consumption of a processor is to perform a task using a continuous frequency by the processor (Figure 2-c). Before moving further, proving the following theorems are necessary:

**Theorem 1:** If $f_1$ and $f_2 (> f_1)$ execute a task in $t_1$ and $t_2$, respectively. Then, $E^{(k)}(f_1, t_1) < E^{(k)}(f_2, t_2)$.

**Proof:**

$$E^{(k)}(f_2, t_2) - E^{(k)}(f_1, t_1) = \\
\left(\alpha f_2^3 + \gamma\right) t_2 + P_{Idle}(T^{(k)} - t_2) - \\
\left(\alpha f_1^3 + \gamma\right) t_1 + P_{Idle}(T^{(k)} - t_1) \\
= \ldots \\
= (f_2 - f_1)\left[\alpha f_1 f_2 (f_2 + f_1) - \gamma + P_{Idle}\right] \\
\geq 0$$

As generally $P_{Idle} > \gamma$, therefore the theorem 1 is proved.

**Theorem 2:** If processor frequency is continues (unrealistic assumption), the optimum energy for k$^{th}$-task is obtained when the task covers the whole task's slack time ($T^{(k)}$).

**Proof:** the result in theorem 1 shows that when a frequency covers the whole slack time it gives the optimum power consumption. Note that this frequency may not exist unless the frequency set is continuous.

Refer to theorem 2, for k$^{th}$-task ($A^{(k)}$), the optimum continuous frequency and its related energy are defined as $f^{(k)}_{opt-cont}$ and $E^{(k)}_{opt-cont}$ and are calculated as [10]:

$$\begin{cases} f^{(k)}_{opt\text{-}cont} = f_N \dfrac{t^{(k)}_{OS}}{T^{(k)}} \\ \\ E^{(k)}_{opt\text{-}cont} = \left(\alpha \left(f^{(k)}_{opt\text{-}cont}\right)^3 + \gamma\right) T^{(k)} \end{cases} \quad (4)$$

In actual systems, however, frequencies must be chosen from a discrete set of frequencies. Also, finishing a task by its deadline may require choosing a frequency that is faster than the optimal frequency. Therefore, the <u>optimal discrete frequency</u> of k$^{th}$-task is the first frequency in the discrete set larger than $f^{(k)}_{opt-cont}$. This discrete frequency and its associated time are $f^{(k)}_{RD}$ and $t^{(k)}_{RD}$, respectively. The algorithm calculating this frequency is referred to as RDVFS for our comparison[7].

## 5.2. Reference Dynamic Voltage-Frequency Scaling (RDVFS)

RDVFS is a simplified version of the algorithm introduced by Kimura et al in [7] for power-scalable high performance clusters supporting DVFS. It reduces energy consumption of processors by selecting the smallest available processor frequency ($f_{RDVFS}$) capable of finishing a task in a given time frame (Figure 2-b). The details of RDVFS algorithm is shown in Figure 3.

---

**RDVFS algorithm:** *slack reclamation by one frequency*
**Input:** *the scheduled tasks on a set of P processors*
1. **for** task $A^{(k)}$ scheduled on processor $P_j$
2.  Compute the optimum continuous frequency ($f^{(k)}_{opt-cont}$) from Eqn.4
3.  Pick the closest higher frequency to $f^{(k)}_{opt-cont}$ in the cpu frequency set, e.g.

$$\left. \begin{array}{l} [f_{max},...,f_n,f_{n-1},...,f_{min}] \\ f_{n-1} < f^{(k)}_{opt-cont} < f_n \end{array} \right\} \Rightarrow f^{(k)}_{RDVFS} = f_n$$

4.  $t^{(k)}_{RDVFS} \leftarrow \dfrac{f^{(k)}_{opt-cont}}{f^{(k)}_{RDVFS}} T^{(k)}$
5.  $E_{RDVFS} \leftarrow f^{(k)}_{RDVFS} t^{(k)}_{RDVFS} + P_{Idle}(T^{(k)} - t^{(k)}_{RDVFS})$
6. **end for**
7. **return** ($f^{(k)}_{RDVFS}$ and $t^{(k)}_{RDVFS}$ for all tasks)

---

Figure 3. RDVFS algorithm

For each task assigned to a processor, $f_{RDVFS}$, which is the first frequency larger than optimal frequency ($f_{opt-cont}$) calculated from Eqn.4, is likely to be the best discrete frequency candidate to execute the task within the given time frame and covering its related slack time. As mentioned before, a major limitation of RDVFS technique is the usage of only one frequency to execute the task.

## 5.3. Maximum- Minimum Frequency for Dynamic Voltage- Frequency Scaling (MMF-DVFS)

Maximum-Minimum Frequency for Dynamic Voltage- Frequency Scaling (MMF-DVFS) technique presented in [10] is similar to RDVFS as both of these approaches use DVFS to reduce energy consumption of scheduled dependent tasks in clusters. Unlike RDVFS algorithm which applies only one frequency to execute a task, MMF-DVFS uses a linear combination of maximum and minimum processor frequencies to achieve the optimal energy consumption regarding to slack time of the task, as shown in figure 2.d. Before explaining further details of MMF-DVFS, proving the following lemma is essential:

**Lemma:** If $f_{DVFS}$ is the appropriate DVFS frequency obtained from RDVFS algorithm with task's energy consumption $E_{DVFS}$; then, there is always a linear combination of the processor's minimum and maximum frequencies with energy consumption less than $E_{DVFS}$.

**Proof:** If $f_N$, $f_1$ and $f_{DVFS}$ are the maximum, minimum and appropriate DVFS processor frequencies extracted from DVFS algorithm, then the lemma indicates that the following non-equation always has a non-zero values for $t_{f_N}$ and $t_{f_1}$ for $k^{th}$ task:

$$\begin{cases} E_{f_N} + E_{f_1} \leq E_{DVFS} \\ t_{f_N} + t_{f_1} \leq T \end{cases} \tag{5}$$

According to Eqn.3 in section 4.2, $E_f \approx Kf^3 t$. By combining this with Eqn.5, we achieve the following equation:

$$\begin{cases} f^3_N * t_{f_N} + f^3_1 * t_{f_1} \leq f^3_{DVFS} * (t_{f_N} + t_{f_1}) \\ t_{f_N} + t_{f_1} \leq T \end{cases} \quad (6)$$

Assuming $t_{f_N} + t_{f_1} = T$, the equation (6) converts to:

$$0 \leq t_{f_N} \leq (\frac{f_{DVFS} - f_1}{f_N - f_{DVFS}}) * t_{f_1}$$

Which indicates that there is always a valid positive $t_{f_N}$ and $t_{f_1}$. The detail of MMF-DVFS algorithm has been shown in Figure 4.

---

**MMF-DVFS algorithm:** *linear combination of maximum and minimum frequencies*
**Input:** *the scheduled tasks on a set of P processors*
1. **for** task $A^{(k)}$ scheduled on processor $P_j$
2.    Calculate amount of time for $f_{max}, f_{min}$ :
   - $t^{(k)}_{f_N} \leftarrow \dfrac{f^{(k)}_{RDVFS} t^{(k)}_{RDVFS} - T^{(k)} f_1}{f_N - f_1}$
   - $t^{(k)}_{f_1} \leftarrow \dfrac{T^{(k)} f_N - f^{(k)}_{RDVFS} t^{(k)}_{RDVFS}}{f_N - f_1}$
3.    $E(k)_{MMF-DVFS} \leftarrow f^3_N t^{(k)}_N + f^3_1 t^{(k)}_1$
4.    **end for**
5. **return** *(the set of $(t^{(k)}_N, t^{(k)}_1)$ for all tasks)*

---

Figure 4. MMF-DVFS algorithm

The algorithm finds the appropriate time portions of the maximum and minimum frequencies to execute each scheduled task. It can be seen from figure 7 that the MMF-DVFS algorithm works the same as RDVFS in the worst case.

In the next section, we present MFS-DVFS algorithm, which uses a linear combination of a variety of processor frequencies instead of two to perform a pre-defined task (Figure 2-e). The new approach is more energy-efficient compared to the other algorithms discussed earlier in this chapter; its energy saving is quite close to the case of using continuous optimum frequency.

## 5.4. Multiple Frequency Selection for Dynamic Voltage-Frequency Scaling (MFS-DVFS)

The RDVFS algorithm decreases a task execution energy by choosing the best processor's speed with respect to the task's idle time [7]. As an example, a set of four tasks scheduled on two processors is shown in Figure.2-a where Figure.2-b, 2-c and 2-d are the results of applying the RDVFS, optimum continuous frequency and MMF-DVFS algorithms on the task, respectively. Figure.2-e also shows the principle of MFS-DVFS algorithm, the proposed algorithm in this chapter. Initially, the task is executed for $t_N^{(k)}$ time units with the highest processor frequency, then its execution frequency is reduced to the second highest value and spends $t_{N-1}^{(k)}$ time units in this frequency. Then, the frequency decreases and task is executed in other frequencies until finishing.

The key idea of MFS-DVFS is to execute tasks using a linear combination of available frequencies so that their slack times are fully filled/covered MFS-DVFS can be defined as finding the best combination of available frequencies $(f_1 < ... < f_N)$ to perform a predefined task with $K$ steps of computation within a predefined time $T$. Therefore, the power consumption minimization of $k^{th}$-task ($A^{(k)}$) in MFS-DVFS algorithm is formulated in an optimization form as follows:

$$\begin{cases} Min: \ E^{(k)} = \sum_{i=1}^{N} t_i^{(k)} (\alpha f_i^3 + \gamma) + P_{Idle}(T^{(k)} - \sum_{i=1}^{N} t_i^{(k)}) \\ s.t. \\ \quad 1. \ \sum_{i=1}^{N} t_i^{(k)} f_i = K^{(k)} \\ \quad 2. \ \sum_{i=1}^{N} t_i^{(k)} \leq T^{(k)} \\ \quad 3. \ t_i^{(k)} \geq 0, \quad for \ i = 1, 2, \cdots, N \end{cases} \qquad (7)$$

The optimization problem in Eqn.7 represents the power consumption problem: how to choose $t_i^{(k)}$ so that the consumed energy of task $A^{(k)}$ minimizes. For executing the task, the processor has to use the same number of tick clocks in both RDVFS and MMF-DVFS algorithms as constraint 1 in Eqn.7. Applying the two mentioned theorems simplifies the optimization problem in Eqn.7 to:

$$\begin{cases} Min: \ E^{(k)} = \sum_{i=1}^{N} t_i^{(k)} (\alpha f_i^3 + \gamma) \\ s.t. \\ \quad 1. \ \sum_{i=1}^{N} t_i^{(k)} f_i = K^{(k)} \\ \quad 2. \ \sum_{i=1}^{N} t_i^{(k)} = T^{(k)} \\ \quad 3. \ t_i^{(k)} \geq 0, \quad for \ i = 1, 2, \cdots, N \end{cases} \qquad (8)$$

To find the best possible values of $t_i^{(k)}$, this optimization algorithm must be applied to all tasks in the scheduling. There are cases that MFS-DVFS cannot improve the power consumption, for example when a task reaches to $f_1$ (the lowest frequency) in the RDVFS algorithm or it has no idle time. Therefore, to improve the speed of MFS-DVFS algorithm, eligible tasks should be extracted before optimization

**Task eligibility:** to simplify the formulation let us just consider 4 discrete values for frequencies (the real processors have normally 4-5 frequencies). In any case, the same procedure can be used for the higher number of frequencies. The problem in Eqn.8 becomes:

$$\begin{cases} Min: \ E^{(k)} = \sum_{i=1}^{4} t_i^{(k)} (\alpha f_i^3 + \gamma) \\ s.t. \\ \quad 1. \ t_1^{(k)} f_1 + t_2^{(k)} f_2 + t_3^{(k)} f_3 + t_4^{(k)} f_4 = K^{(k)} \\ \quad 2. \ t_1^{(k)} + t_2^{(k)} + t_3^{(k)} + t_4^{(k)} = T^{(k)} \\ \quad 3. \ t_i^{(k)} \geq 0, \quad for \quad i = 1, 2, \cdots, 4 \end{cases}$$

Merging constraints 2 and 3 results in:

$$\begin{cases} t_1^{(k)} = \dfrac{T^{(k)} f_2 - K^{(k)}}{f_2 - f_1} - t_3^{(k)} \dfrac{f_2 - f_3}{f_2 - f_1} - t_4^{(k)} \dfrac{f_2 - f_4}{f_2 - f_1} \\ t_2^{(k)} = \dfrac{K^{(k)} - T^{(k)} f_1}{f_2 - f_1} - t_3^{(k)} \dfrac{f_3 - f_1}{f_2 - f_1} - t_4^{(k)} \dfrac{f_4 - f_1}{f_2 - f_1} \end{cases}$$

Therefore, the power consumption function changes to

$$E^{(k)} = a_0^{(k)} + a_1^{(k)} t_3^{(k)} + a_2^{(k)} t_4^{(k)} \tag{9}$$

Where

$$a_0^{(k)} = (\alpha f_1^3 + \gamma)\frac{T^{(k)} f_2 - K^{(k)}}{f_2 - f_1} + (\alpha f_2^3 + \gamma)\frac{K^{(k)} - T^{(k)} f_1}{f_2 - f_1}$$

$$a_1^{(k)} = (\alpha f_3^3 + \gamma) + (\alpha f_1^3 + \gamma)\frac{f_3 - f_2}{f_2 - f_1} - (\alpha f_2^3 + \gamma)\frac{f_3 - f_1}{f_2 - f_1} \quad (10)$$

$$a_2^{(k)} = (\alpha f_4^3 + \gamma) + (\alpha f_1^3 + \gamma)\frac{f_4 - f_2}{f_2 - f_1} - (\alpha f_2^3 + \gamma)\frac{f_4 - f_1}{f_2 - f_1}$$

To guarantee achieving less energy consumption using MFS-DVFS algorithm, the following condition should be satisfied.

$$a_0^{(k)} + a_1^{(k)} t_3^{(k)} + a_2^{(k)} t_4^{(k)} < E_{RD}^{(k)} \quad (11)$$

$a_0^{(k)} + a_1^{(k)} t_3^{(k)} + a_2^{(k)} t_4^{(k)}$ shows a 3-dimensional surface and the search region is where it satisfies the three following constraints: (1) $t_3^{(k)} \geq 0$, (2) $t_4^{(k)} \geq 0$ and (3) $E_{RD}^{(k)} > 0$. The first two constraints in Eqn.11 are also considered by optimization in Eqn.8. The only one that specifies the search region is constraint 3. If a task satisfies this recent constraint, then it can be concluded that there is a valid search region for this task where MFS-DVFS gives better result than RDVFS. Then linear programming explores this search region to find out the best suitable frequencies and their associated times. The detail of MFS-DVFS algorithm has been shown in Figure 5.

> ***MFS-DVFS algorithm:*** *linear combination of frequencies*
> ***Input:*** *the scheduled tasks on a set of P processors*
> 1. ***For*** *task $A^{(k)}$ scheduled on processor $P_j$*
> 2.   *Apply RDVFS algorithm on this task*
> 3.   ***if*** $E_{RD}^{(k)} > 0$ *for this task **then***
>        *- this task is eligible for MFS-DVFS*
>        *- Solve optimization problem in Eqn6 by linear programming*
>      ***else***
>        *RDVFS is the optimal result*
> 4.   ***end if***
> 5. ***end for***
> 6. ***return*** *(the voltages and frequencies of optimal execution)*

Figure 5. MFS-DVFS algorithm

# 6. Experimental Results

In this section we present the results of energy consumption obtained from simulating our MFS-DVFS algorithm in comparison with RDVFS, MMF-DVFS and optimum continuous frequency. In order to compare the algorithms, the following schedulers were used with different number of processors: (1) list scheduling, (2) list scheduling with Longest Processing Time first (LPT) and (3) list scheduling with Shortest Processing Time first (SPT).

The simulations were carried out using the simulator we developed as a part of this study.

## 6.1. Simulation Settings

We use the voltage/frequency setting of two real processors in our simulations: Transmeta Crusoe [7] and Intel Xscale [22]. Table 1 shows the voltage/frequency and the related power consumption of these processors following with the convex models of each processor. These models use least-square curve fitting to fit a convex function $(\alpha f^3 + \gamma)$ on the frequency-power of two real processors, as shown in Figure 6.

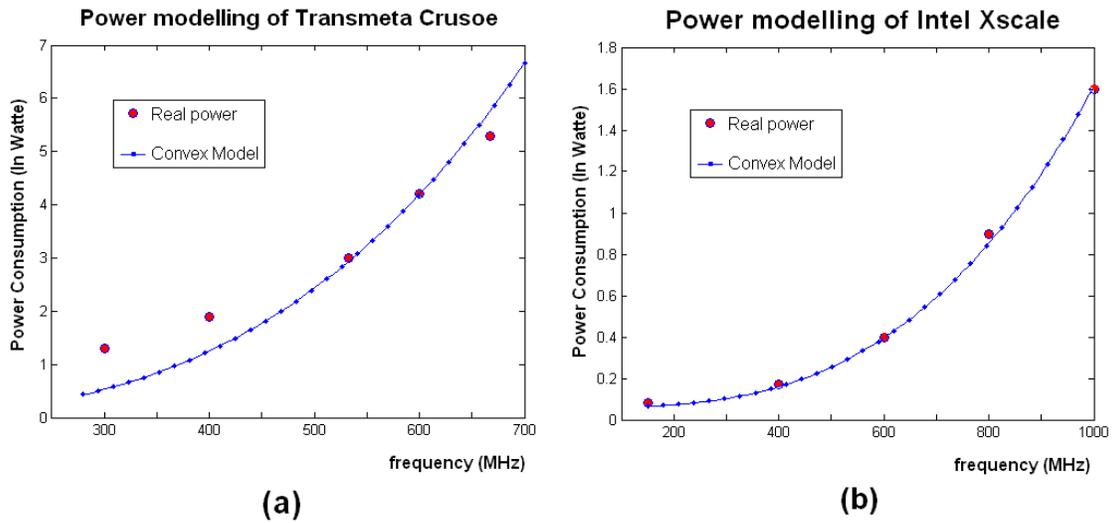

Figure 6. The least-square modelling of (a) Transmeta Crusoe, and (b) Intel Xscale processors

We evaluated the performance of MFS-DVFS with two sets of task graphs: randomly generated and real-word parallel applications. The two real world applications used in our experiments were LU decomposition and Gauss-Jordan with DAGs extracted from [19]. We applied a large number of variations in the number of processors and tasks for each application in our simulations. The random task graph set consisted of 1500 graphs with five graph sizes of 100, 200, 300, 400 and 500 nodes, together with three different schedulers on five sets of 2, 4, 8, 16 and 32 processors.

**TABLE 1.** The voltage/frequency setting of two real processors in the experiments with their power consumption and convex models

| Level | *Transmeta Crusoe* | | |
|---|---|---|---|
| | Frequency (MHz) | Voltage (V) | Power (W) |
| 0 | 667 | 1.6 | 5.3 |
| 1 | 600 | 1.5 | 4.2 |
| 2 | 533 | 1.35 | 3.0 |
| 3 | 400 | 1.225 | 1.9 |
| 4 | 300 | 1.2 | 1.3 |
| Convex model | $P = 1.94 \times 10^{-5} \left( \dfrac{f}{10^6} \right)^3 + 4.44$ mW | | |
| Level | *Intel Xscale* | | |
| | Frequency (MHz) | Voltage (V) | Power (W) |
| 0 | 1000 | 1.8 | 1.6 |
| 1 | 800 | 1.6 | 0.9 |
| 2 | 600 | 1.3 | 0.4 |
| 3 | 400 | 1 | 0.17 |
| 4 | 150 | 0.75 | 0.08 |
| Convex model | $P = 1.55 \times 10^{-6} \left( \dfrac{f}{10^6} \right)^3 + 60$ mW | | |

These task graphs have different number of tasks, task distributions, communication costs and task dependencies. The execution cycle of these randomly generated tasks varied from 5-10 million cycles from a uniform distribution, respectively. We used 150 real-world application task graphs based on LU decomposition algorithm in our experiments. For the real-application graph, the same number of task graphs –ranging from 100 to 500 tasks– with three schedulers and on five sets of processors were investigated.

## 6.2. Results

The simulation results of normalized energy consumption for all DAGs (Figures. 7 and 8) are shown in table 2. This table clearly denotes the superior performance of MFS-DVFS scheduling compared to the other approaches in all cases. Figure 8 depicts that although the efficiency of all algorithms including MFS-DVFS in saving energy in LU decomposition is significant, these algorithms have less performance on Gauss-Jordan tasks. For a deeper examination of this behaviour, a sample three level Gauss-Jordan application job scheduling on three processors has been shown in Figure 9. As explained before, since there is no idle time among tasks in Gauss-Jordan graphs applications, none of these algorithms can efficiently reduce energy consumption.

An interesting issue for further investigation is the relationship between energy consumption and the number of processors in our experiments. Increasing the number of processors expedites the processing time and consequently reduces the makespan; however, as a drawback, it also increases the system slack time. Figure 10 addresses this issue and illustrates the percentage of overall energy saving of the system on the number of processors for random and LU decomposition task graphs. The graphs in this figure reveal the fact that increasing the number of processors results in saving more energy.

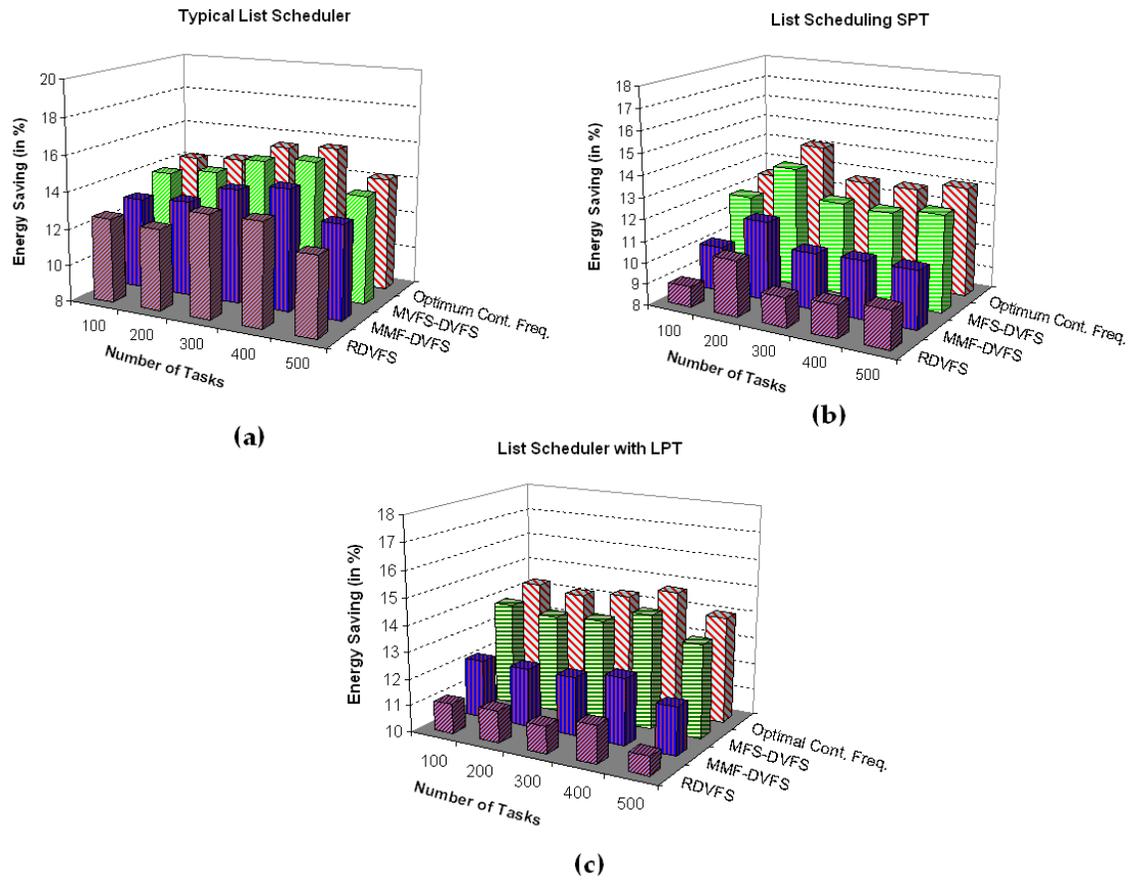

Figure 7. The normalized energy consumption on the number of tasks: (a) The typical list scheduler (b) The list scheduler with Longest Processing Time first (LPT) and (c) The list scheduler with Shortest Processing Time first (SPT).

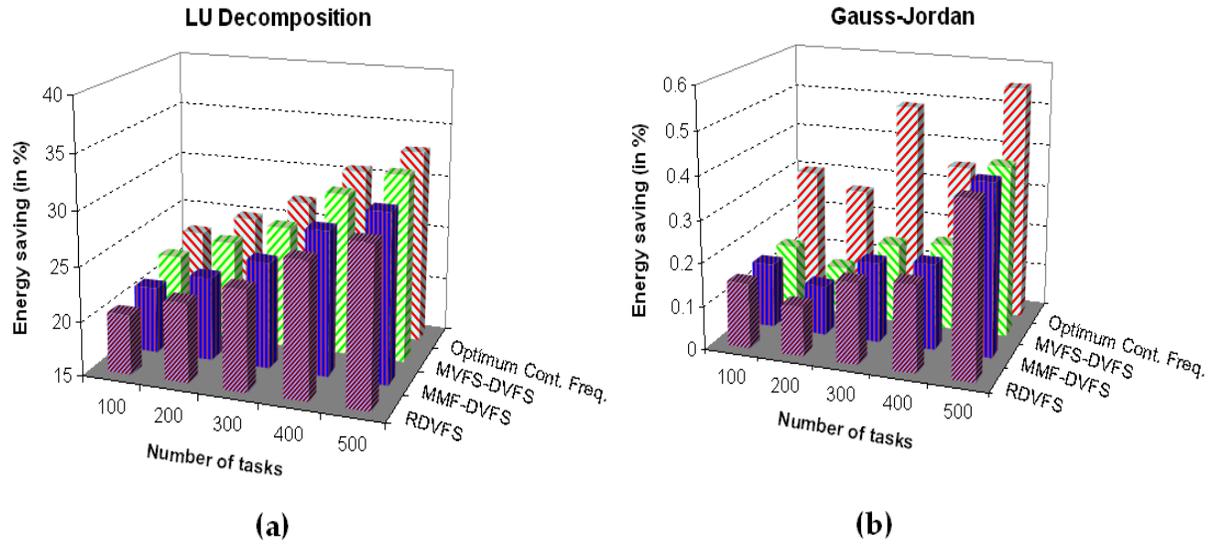

Figure 8. The normalized energy consumption of MFS-DVFS and other algorithms on the number of tasks for two real-world applications: (a) LU decomposition, (b) Gauss-Jordan.

Table 2. The energy saving percentage of MFS-DVFS and other algorithms on 1800 random and real task graphs.

| Experiment | Random tasks | Gauss-Jordan | LU-decomposition |
|---|---|---|---|
| *RDVFS* | 13.00% | 0.1% | 24.8% |
| *MMF-DVFS* | 13.50% | 0.11% | 25.5% |
| *MVFS-DVFS* | 14.40% | 0.11% | 27.0% |
| *Optimal Continuous Frequency* | 14.84% | 0.14% | 27.81% |

The major limitation on most DVFS-based algorithms working with one frequency (such as the RDVFS algorithm) is that the frequency combinations are fixed. Those algorithms work better when the processor can run at any arbitrary set of frequencies. However, due to technological issues, the number of valid frequencies is limited so that these algorithms have to choose the most appropriate frequency among a set of frequencies defined by DVFS. According to the fix number of tick cycles for a task (constraint 1 in Eqn.8) the relation among $t_{RD}^{(k)}, f_{RD}^{(k)}, f_N$ and $t_{OS}^{(k)}$ for task $A^{(k)}$ is:

$$t_{RD}^{(k)} = \frac{f_{RD}^{(k)}}{f_N} t_{OS}^{(k)}$$

It is shown that although $t_{RD}^{(k)}$ is a continuous variable, it cannot accept all values; therefore the slack time of tasks cannot be minimized. However, in MFS-DVFS algorithm, the relation between those variables is

$$f_{RD}^{(k)} t_{RD}^{(k)} = f_1 t_{1_{RD}}^{(k)} + f_2 t_{2_{RD}}^{(k)} + \ldots + f_N t_{N_{RD}}^{(k)}$$

which is one equation with more than one variables $(t_{1_{RD}}^{(k)}, \ldots, t_N^{(k)})$ and might have many eligible results; thus, appropriate values of these variables, with regard to the task conditions, can minimize the slack time and/or reduce energy consumption.

An overhead with MFS-DVFS and MMF-DVFS is the transition time of switching from the one frequency to another one. An almost true assumption is that the overhead of transition times is relatively much less than the execution times of tasks; therefore the transition times overhead can be neglected in calculations. In our experiments, the tasks with T at least 20 times more than transition time is considered for the MFS-DVFS algorithm.

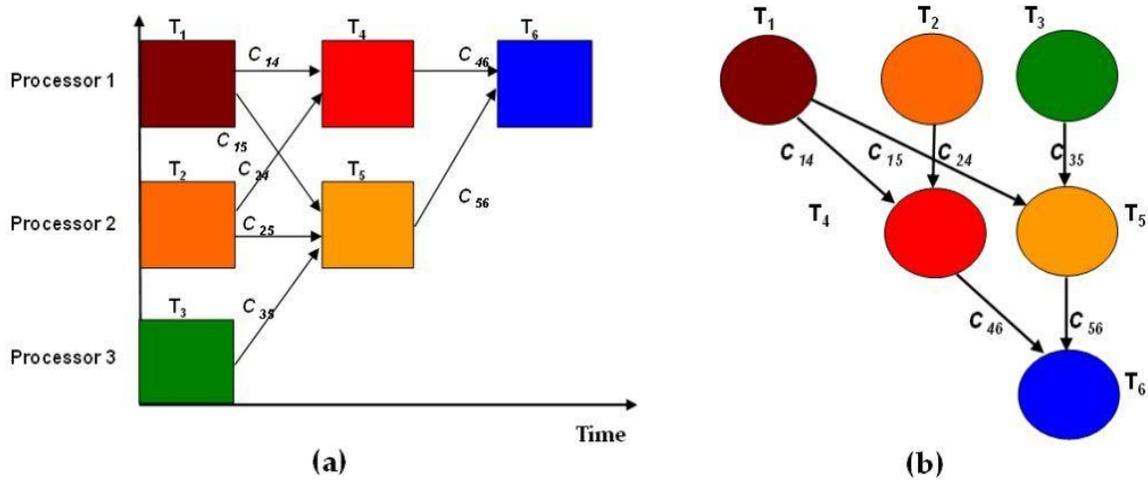

Figure 9. Gauss-Jordan task graph: (a) a sample scheduling of a three level Gauss-Jordan task graph on three processors, (b) a Gauss-Jordan DAG for three levels. The communication costs ($Cij$) are equal to 10 time units for all $i$ and $j$.

## 7. Conclusion

Since most traditional static task scheduling algorithms in HPCS do not consider power management, we addressed the energy issue with task scheduling and presented the MFS-DVFS algorithm. Our algorithm adopted the DVFS technique, a recent advance in processor design, to reduce energy consumption.

In this chapter, we studied existing DVFS-based approaches to cover idle time and in particular, using a linear combination of more than one frequency to reduce energy consumption on processors. First, we noticed the energy model in DVS-enabled processors. Then, we formulated our algorithm (MFS-DVFS) as an optimization problem of all frequencies for each task and then solved it to find the suitable time portions. Simulation results of 1500 randomly generated task graphs and 300 real world application task graphs showed the effectiveness of the MFS-DVFS algorithm compared with other algorithms.

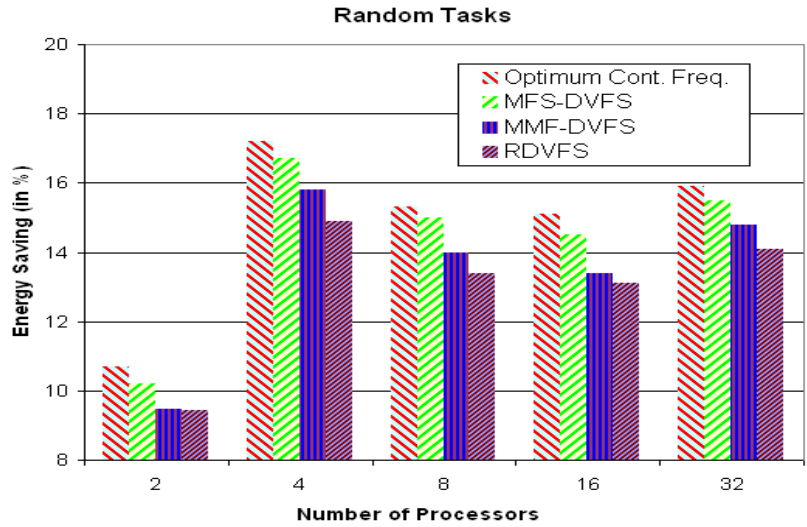

(a)

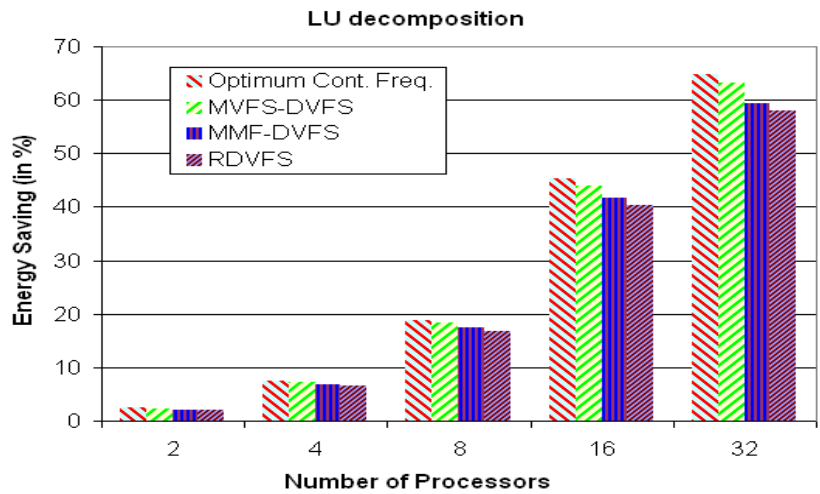

(b)

Figure 10. The comparison between the percentages of energy saving in MFS-DVFS with other algorithms on the number of processors: (a) 1500 randomly generated task graphs, (b) 300 LU decomposition task graphs.

## 8. Acknowledgment

The work reported in this chapter is in part supported by National ICT Australia (NICTA). Professor A.Y. Zomaya's work is supported by an Australian Research Council Grant LP0884070.

## REFERENCE

[1] N. Kamyabpour and D. B. Hoang, "A hierarchy energy driven architecture for wireless sensor networks," presented at the 24th IEEE International Conference on Advanced Information Networking and Applications (AINA-2010), Perth, Australia, 2010.

[2] N. Kamyabpour and D. B. Hoang, "A Task Based Sensor-Centeric Model for overall Energy Consumption," Computing Research Repository(*CoRR*), 2012.

[3] K. Almiani, S. Selvakennedy, and A. Viglas, "RMC: An Energy-Aware Cross-Layer Data-Gathering Protocol for Wireless Sensor Networks," presented at the 22nd International Conference on Advanced Information Networking and Applications (AINA), GinoWan, Okinawa, Japan, 2008.

[4] K. Almiani, A. Viglas, and L. Libman, "Energy-efficient data gathering with tour length-constrained mobile elements in wireless sensor networks," presented at the The 35th Annual IEEE Conference on Local Computer Networks (LCN), Denver, Colorado, USA, 2010.

[5] J. Zhuo and C. Chakrabarti, "Energy-efficient dynamic task scheduling algorithms for DVS systems," *ACM Trans. Embed. Comput. Syst.,* vol. 7, pp. 1-25, 2008.

[6] R. Ge, X. Feng, and K. W. Cameron, "Performance-constrained Distributed DVS Scheduling for Scientific Applications on Power-aware Clusters," presented at the Proceedings of the 2005 ACM/IEEE conference on Supercomputing, Seattle, WA, USA, 2005.

[7] H. Kimura, M. Sato, Y. Hotta, T. Boku, and D. Takahashi, "Emprical study on Reducing Energy of Parallel Programs using Slack Reclamation by DVFS in a

Power-scalable High Performance Cluster," *in 2006 IEEE International Conference on Cluster Computing,*, Barcelona, Spain, 2006, pp. 1-10.

[8]     R. Xiaojun, Q. Xiao, Z. Ziliang, K. Bellam, and M. Nijim, "An Energy-Efficient Scheduling Algorithm Using Dynamic Voltage Scaling for Parallel Applications on Clusters," in *Computer Communications and Networks, 2007. ICCCN 2007. Proceedings of 16th International Conference on*, Honolulu, Hawaii, USA, 2007, pp. 735-740.

[9]     Y. C. Lee and A. Y. Zomaya, "Minimizing Energy Consumption for Precedence-Constrained Applications Using Dynamic Voltage Scaling," presented at the Proceedings of the 2009 9th IEEE/ACM International Symposium on Cluster Computing and the Grid (CCGrid), Shanghai, China, 2009.

[10]    N. B. Rizvandi, J. Taheri, A. Y. Zomaya, and Y. C. Lee, "Linear Combinations of DVFS-enabled Processor Frequencies to Modify the Energy-Aware Scheduling Algorithms," presented at the Proceedings of the 2010 10th IEEE/ACM International Symposium on Cluster, Cloud and Grid Computing (CCGrid), Melbourne, Australia, May 17-20, 2010.

[11]    A. Berl, E. Gelenbe, M. d. Girolamo, G. Giuliani, H. d. Meer, M. Q. Dang, and K. Pentikousis, "Energy-Efficient Cloud Computing," *The Computer Journal,* 2009.

[12]    N. Kappiah, V. W. Freeh, and D. K. Lowenthal, "Just In Time Dynamic Voltage Scaling: Exploiting Inter-Node Slack to Save Energy in MPI Programs," presented at the Proceedings of the 2005 ACM/IEEE conference on Supercomputing, Seattle, USA, 2005.

[13]    Z. Zhu and X. Zhang, "Look-Ahead Architecture Adaptation to Reduce Processor Power Consumption," *IEEE Micro,* vol. 25, pp. 10-19, 2005.

[14]    D. Yang, K. Malkowski, P. Raghavan, and M. Kandemir, "Towards energy efficient scaling of scientific codes," in *Parallel and Distributed Processing, 2008. IPDPS 2008. IEEE International Symposium on*, Miami, USA, 2008, pp. 1-8.

[15]    A. Molnos and K. Goossens, "Conservative Dynamic Energy Management for Real-Time Dataflow Applications Mapped on Multiple Processors," presented at the 12th Euromicro Conference on Digital System Design, Architectures, Methods and Tools, Patras, Greece, 2009.

[16]    Y. Hotta, M. Sato, H. Kimura, S. Matsuoka, T. Boku, and D. Takahashi, "Profile-based optimization of power performance by using dynamic voltage scaling on a pc cluster," presented at the IEEE International Symposium on Parallel and Distributed Processing (IPDPS), Isle of Rhodes, Greece, 2006.

[17]    R. Springer, D. K. Lowenthal, B. Rountree, and V. W. Freeh, "Minimizing execution time in MPI programs on an energy-constrained, power-scalable

cluster," presented at the Proceedings of the eleventh ACM SIGPLAN symposium on Principles and practice of parallel programming, New York, New York, USA, 2006.

[18] B. Rountree, D. K. Lowenthal, S. Funk, V. W. Freeh, B. R. de Supinski, and M. Schulz, "Bounding energy consumption in large-scale MPI programs," in *Supercomputing, 2007. SC '07. Proceedings of the 2007 ACM/IEEE Conference on*, 2007, Reno, Nevada, USA, pp. 1-9.

[19] T. Simunic, L. Benini, A. Acquaviva, P. Glynn, and G. D. Micheli, "Dynamic voltage scaling and power management for portable systems," presented at the Proceedings of the 38th annual Design Automation Conference, Las Vegas, Nevada, USA, 2001.

[20] P. d. Langen and B. Juurlink, "Trade-Offs Between Voltage Scaling and Processor Shutdown for Low-Energy Embedded Multiprocessors," presented at the Embedded Computer Systems: Architectures, Modeling, and Simulation, Samos, Greece, 2007.

[21] J.-J. Chen, C.-Y. Yang, T.-W. Kuo, and C.-S. Shih, "Energy-Efficient Real-Time Task Scheduling in Multiprocessor DVS Systems," presented at the Proceedings of the 2007 Asia and South Pacific Design Automation Conference, Yokohama, Japan, 2007.

[22] C. Xian and Y.-H. Lu, "Dynamic voltage scaling for multitasking real-time systems with uncertain execution time," presented at the Proceedings of the 16th ACM Great Lakes symposium on VLSI, Philadelphia, PA, USA, 2006.